\documentclass[letterpaper,11pt]{report}

\pdfoutput=1

\usepackage{amsmath, amsthm, amssymb}
\usepackage[dissertation]{USCthesis}
\usepackage{url}  
\usepackage{ifthen}  
\usepackage{multicol}   
\usepackage[utf8]{inputenc} 
\usepackage{amsthm}
\usepackage{graphics}
\usepackage{graphicx}

\usepackage{comment}

\usepackage{algorithm}
\usepackage{algorithmic}

\usepackage[numbers]{natbib}

\theoremstyle{plain}

\theoremstyle{definition}

\begin{document}
\title{\doublespacing{AUTOMATIC TRACKING OF PROTEIN VESICLES}}
\author{Min Xu}
\submitdate{May 2009}

\universityname{UNIVERSITY OF SOUTHERN CALIFORNIA}
\schoolname{GRADUATE SCHOOL} \degree{MASTER OF ARTS}
\majorfield{APPLIED MATHEMATICS}

\committee{Sergey V. Lototsky & (Chair person)\\*
               Remigijus Mikulevicius \\*
               Donald B. Arnold & (Outside Member)}

\begin{preface}
\prefacesection{Dedication}
\begin{center}
{To Luge and Shiyou}
\end{center}

\prefacesection{Acknowledgements}

I realize this thesis with the help of several wonderful and kind people. First of all I would like to thank my advisor, Professor Sergey Lototsky for his great help, patience and indispensable advises. Also I would like to thank Professor Donald Arnold and Mr. Sarmad Al-Bassam. My interest on designing methods of tracking protein vesicles is primarily motivated from their research. Finally, I would like to thank Professor Remigijus Mikulevicius for reviewing my thesis and attending my presentation.

\tableofcontents
\listoffigures

\prefacesection{Abstract}
With the advance of fluorescence imaging technologies, recently cell biologists are able to record the movement of protein vesicles within a living cell. Automatic tracking of the movements of these vesicles become key for qualitative analysis of dynamics of theses vesicles. In this thesis, we formulate such tracking problem as video object tracking problem, and design a dynamic programming method for tracking single object. Our experiments on simulation data show that the method can identify a track with high accuracy which is robust to the choose of tracking parameters and presence of high level noise. We then extend this method to the tracking multiple objects using the track elimination strategy. In multiple object tracking, the above approach often fails to correctly identify a track when two tracks cross. We solve this problem by incorporating the Kalman filter into the dynamic programming framework. Our experiments on simulated data show that the tracking accuracy is significantly improved.

\end{preface}

\chapter{Introduction} \label{chp:introduction}

\begin{figure} [tph]
\centering
\includegraphics[width=1.0\columnwidth] {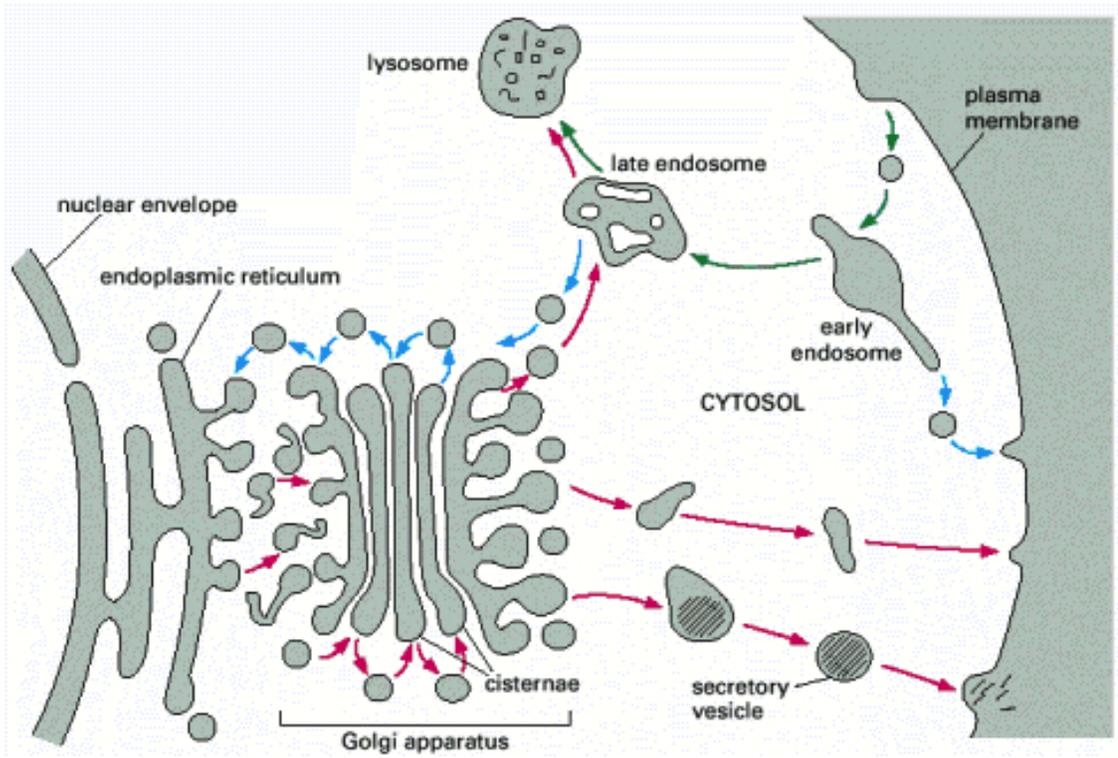}
\caption{The intracellular compartments of the eucaryotic cell involved in the biosynthetic secretory pathways. }
\label{fig:biosynthetic_pathway}
\end{figure}

Every cell must communicate with the world around it. Eucaryotic cells (i.e. cells of animals, plants, and fungi) have internal membrane system that allows them to regulate the delivery of newly synthesized proteins to the cell exterior. The biosynthetic-secretory pathway allows the cell to modify the molecules it produces in a series of steps, store them until needed, and then deliver them to the exterior. Such delivery is through protein vesicles, which are small bubbles of liquid within a cell. Figure~\ref{fig:biosynthetic_pathway} conceptualize typical biosynthetic-secretory pathways in a cell. In the figure, each compartment encloses a space, called a lumen, that is topologically equivalent to the outside of the cell, and all compartments shown communicate with one another and the outside of the cell by means of transport vesicles. In the biosynthetic-secretory pathway (red arrows) protein molecules are transported from the endoplasmic reticulum (ER) to the plasma membrane or (via late endosomes) to lysosomes. Some molecules are retrieved from the late endosome and returned to the Golgi apparatus, and some are retrieved from the Golgi apparatus and returned to the ER. The figure and caption are adapted from ~\cite{alberts2002mbc}.

Fluorescence microscopy is a main tool to study the biosynthetic-secretory processes. A cell is normally optical transparent. To visualize the molecules of a protein of interest, they can be labeled using fluorescence dye. When excited using light of a particular wavelength, the dye can emit light of another wavelength that can be detected. Thus, the location of the protein molecules in a cell can be identified. In recent years, the resolution of location identification is highly increased through the advance of confocal microscopy techniques. In addition, the discovery of fluorescent proteins enables biologists to observe the dynamics of proteins in individual  \emph{living} cells. So far, little has been done on automatic analysis of the dynamics of protein molecules. Here we focus on tracking movement of protein vesicles from microscopy image sequence (i.e. a video).

\begin{figure} [tph]
\centering
\includegraphics[width=0.32\columnwidth] {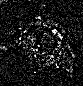}
\includegraphics[width=0.32\columnwidth] {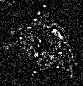}
\includegraphics[width=0.32\columnwidth] {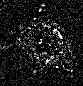}
\caption{Three consecutive fluorescence microscopy images showing protein vesicles. }
\label{fig:vesicles_substract}
\end{figure}

Suppose microscopy images are grey scale images. After removing static structure in a image sequence, protein vesicle becomes a spot like object that is relative brighter (i.e. has a higher value) than dark background. Figure ~\ref{fig:vesicles_substract} shows such an image. In the figure, majority of the static cell structure are suppressed. video obtained from ~\cite{kockx2007sam}. There are two main challenges in such kind of tracking : 1) high level of noise in images and 2) tracking of \emph{multiple} objects.

\chapter{Literature review} \label{chp:literature}

\section{Object tracking methods}
Object tracking method is an important first step of automatic analyzing cellular dynamics. Object tracking is originally developed in the field of computer vision. Basically, tracking can be defined as the problem of estimating the trajectory of an object in the image plane as the object moves around a scene. In practice, there are many difficulties to successful building of a tracker algorithm. The difficulties related to protein vesicle tracking are 1) noise in images, 2) complex object motion, 3) partial and full object occlusions. The above problems sometimes can be simplified by incorporating prior knowledge of the objects. However, so far little is known about the characteristics of molecular dynamics in cells.

Numerous methods for object tracking have been proposed in the field of computer vision. They mainly differ in the following aspects~\cite{yilmaz2006ots}: 1) object representation 2) image features used 3) modeling of motion, appearance, and shape of the object. These methods address the above aspects according to the context/environment in which the tracking is performed and the tracking information needed for subsequent analysis.

\subsection{Object representation}

The shape of an object can be represented in different ways: 1) points, i.e., the centroid ~\cite{veenman2001rmc} or a finite set of points ~\cite{serby2004pot}. 2) Primitive geometric shapes (eg ellipse, rectangle etc) ~\cite{comaniciu2003kbo}. 3) object silhouette and contour ~\cite{yilmaz2004cbo}. 4) Articulated shape models, i.e., shapes held together with joints. 5) Skeleton  ~\cite{ballard1982cv}. Since in our case, the protein vesicles only occupy small regions in an image, we normally represent these vesicles as points.

In combining shape representations, the are a number of ways to represent the appearance features of objects. Following are a number of appearance features: 1) probability densities, can be either parametric ~\cite{zhu1996rcu} or nonparametric ~\cite{elgammal2002baf}; 2) active appearance models ~\cite{edwards1998ifi}, which associate landmarks with feature vectors of color, texture, etc; 3) multiview appearance models, which are generated from subspaces of given views using techniques like Principal Component Analysis (PCA) and Independent Component Analysis (ICA) ~\cite{moghaddam1997pvl}.

\subsection{Object detection}
A tracking method needs to be able to determine the existence of an object in every frame or in the frame that the object first appears in the video. Some detection methods use the temporal information computed form a sequence of frames to reduce false detections. Following are several types of common methods used for object detection.

Point Detectors: point detectors are used to identify points that belong to an object, based on local context. Commonly used point detectors are: Moravec's interest operator, Harris interest point detector, KLT detector, and SIFT detector ~\cite{mikolajczyk2002aii}.

Background Subtraction: object detection can also be achieved by constructing a background model and then finding deviations from the model. A significant change in an image region from the background model would indicate a moving object. This process is called background substraction. Due to the increasing abilities to efficiently model complex background, most of recent tracking methods for fixed cameras use background subtraction methods to detect regions of interest (for example ~\cite{haritaoglu2000wrt}, ~\cite{collins2001acm}).

Segmentation: image segmentation algorithms is used to partition the image into perceptually similar regions. A segmentation algorithm normally defines criteria for a good partition and provides a method for efficient partitioning. Several types of segmentation methods have been designed for tracking purpose: 1) mean shift clustering ~\cite{comaniciu1999msa}, which performs clustering in the joint spatial and color space; 2) graph cutting, which converts the image into a graph, and uses min-cut algorithm to find disjoint regions ~\cite{shi2000nca}; 3) active contours, which evolves a closed contour to the objects' boundary, evaluated by certain energy function (for example ~\cite{kass1988sac}).

Supervised Learning: in this approach, object detection is formulated as a classification problem. The learning algorithm is used to generate models from the features of the objects in the training video where the objects are known. Then these models are used to predict the existence of objects in new video data. When a set of image features are properly chosen, a number of different classification methods can be used, such as neural networks ~\cite{rowley1998nnb}, adaptive boosting ~\cite{viola2005dpu}, decision trees ~\cite{grewe1995ilm}, and support vector machines ~\cite{papageorgiou1998gfo}.

\subsection{Object tracking}
Object tracking is used to generate the trajectory of an object over time by locating its position in every frame of the video. There are two tasks in object tracking: detecting the object and establishing association of objects between frames. These two tasks can be performed separately or jointly. For different object representation, different types of tracking methods are developed. They fall into three categories: 1) point tracking, which represents objects as points and only estimates the object's position in each frame; 2) kernel tracking, which uses object shape and appearance and considers not only translation but also rotation of objects; 3) silhouette tracking, which uses template matching to identify objects in each frame. Because in our project, the protein vesicles are mainly represented as points, in this section, we will only describe point tracking methods.

In point tracking, tracking can be formulated as the association of detected objects represented by points across frames. In general, there are two types of methods to associate points: deterministic and statistical methods.

Deterministic methods usually define a cost of associating each object in frame $t-1$ to a single object in frame $t$ using a set of motion constraints. Minimization of the association cost is formulated as a combinatorial optimization problem. Optimal assignment methods are developed to obtain the best one-to-one association among all possible associations (for example, Hungarian algorithm ~\cite{kuhn1955hms}). The association cost is usually defined by using a combination of the following constraints: 1) object displacement between frames; 2) maximum velocity; 3) small velocity change; 4) common motion, which requires the velocity of objects in a small neighborhood to be similar; 5) rigidity, which assumes that objects are rigid.

Here are a number of such methods. Sethi and Jain ~\cite{sethi1987ftf} proposed a greedy method to solve the association problem. It is based on the proximity and rigidity constraints applied on two consecutive frames. Salari and Sethi ~\cite{salari1990fpc} designed a method that establishes association for the detected points and then extend the tracking of the missing objects by adding hypothetical points. From their work, Veenman et al. ~\cite{veenman2001rmc} further added the common motion constraint. Shafique and Shah ~\cite{shafique2005nga} uses the temporal coherency of speed and position in multiple frames as constrain. In their approach, the association problem is converted to finding the best unique path for each point on a graph.

Compared to deterministic methods, statistical association methods are used to reduce the effect of noise in the video and the perturbation of movements in objects by incorporating the randomness into model. They use the state space approach to model the object properties such as position, velocity, and acceleration.

For single object tracking, a typical statistical association method is Kalman filter ~\cite{broida1986eom}, which assumes the transition of system states is linear and the noise is Gaussian. We will use Kalman filter to enhance our tracking method. One limitation of Kalman filter is the assumption of Gaussian distribution of the state variables (see Method section for details). Particle filtering ~\cite{mackay1998imc} has been used to reduce the above limitation through model estimation by importance sampling.

Multiobject association and state estimation are often carried out statistically. When tracking multiple objects using Kalman or particle filters, the association problem needs to be solved before these filters can be applied. There are two popular methods for data association: Joint Probability Data Association Filtering (JPDAF) ~\cite{chang1991sre} and Multiple Hypothesis Tracking (MHT) ~\cite{reid1979atm}. JPDAF extends Kalman filter by replacing its innovation of single track with a sum of innovations of multiple tracks weighted by the posterior probability that a measurement is associated with that track. On the other hand, MHT iteratively improves associations. In each iteration, the algorithm starts from a set of current track hypotheses, in form of collections of disjoint tracks. For each hypothesis, the algorithm predicts each object's position in the next frame. By comparing the predictions with actual measurements, associations are established for each hypothesis and a set of new hypotheses are formed for next iteration.

\section{Tracking methods applied to cellular dynamics}

The field of tracking of molecular dynamics in cells is relatively new. So far only a small number of methods are developed for this purpose. They are summarized as follows.

Sbalzarini et al. ~\cite{sbalzarini2005fpt} proposed a tracking method that consists of two steps: feature detection and trajectory linking. In this approach, proteins are represented as points. The feature detection step mainly consists of detection of refinement of points according to local maxima values. Given the detected candidate locations, trajectory linking then associates the points between each of two adjacent frames. 

Godinez et al. ~\cite{godinez2007tvp} designed a method for tracking virus particles. The method consists of virus particle detection and association. For the detection step, they use Laplacian-of-Gaussian filtering for detecting spots. Laplacian-of-Gaussian is a image filtering technique that applies Gaussian blur and Laplacian operator to a image. They then use Gaussian fitting to enhance spots. For the association step, they employed a smooth motion and nearest neighbor constraints to link detected particles between frames.

Sage et al. ~\cite{sage2005ati} proposed a dynamic programming approach for tracking the fluorescent markers attached in a single chromosome in a cell. Basically, given a discrete scalar field, dynamic programming is a computational technique that can be used to find a curve such that integration along this curve would achieve optimal value. The advantage of such approach is that object detection step is not required. We will describe this approach in detail in the next section.

The above approaches are deterministic. A few stochastic tracking approaches also appeared recently. For example, Yoon et al. ~\cite{yoon2008bii} proposed to use particle filter to track a single molecule. Simply speaking, when the states of objects are modeled as a Markov Chain, particle filter obtains optimal Bayesian estimation of states given noisy observations over time. Also using particle filter, Smal et al. ~\cite{smal2007rbm} designed a method for tracking microtubles in a cell. In their approach, microtubes are modeled using Gaussian functions for the detection. After detecting the molecules, particle filter is used for estimation of tracks.

\chapter{Tracking of single vesicle} \label{chp:single}
\section{Problem formulation}

In this section, we study the tracking of a single vesicle in a video. We solve single vesicle tracking problem through an optimization approach given a  sequence of $n$-dimensional images (normally $n=2$). Denote $\mathbb{X} \subset \mathbf{R}^n$ as the set of all possible locations, which is identical for all images. Let $f(\mathbf{x}, t)$ be the intensity of location $\mathbf{x} \in \mathbf{X}$ of the image at time $t$, we want to find a track $\mathbf{x}_t, \ t = 1,\ldots,T$ such that the following score function is maximized:

\begin{equation}
s_T = \sum_{t=1}^T f(\mathbf{x}_t, t) - w (\sum_{t=2}^T \| \mathbf{x}_t - \mathbf{x}_{t-1} \| ))
\label{eqn:score_function}
\end{equation}

Intuitively, this score function tends to be high when the intensity along the track is high and displacement is low. These two factors are balanced using a with a weight $w$.

\section{A dynamical programming approach}

Such optimization problem can be solved using dynamic programming technique~\cite{corman2001ia}. The above problem can be decomposed into subproblems, and optimal solutions of subproblems can be used to find the optimal solutions of the overall problem. Formally, let $s_{t}(\mathbf{x})$ be the maximum score of all tracks of length $t$ that ends up at position $\mathbf{x}$. That is:

\begin{equation}
s_{t}(\mathbf{x}) \triangleq max_{\mathbf{x}_1, \ldots, \mathbf{x}_{t-1}} \,\,
\{ f(\mathbf{x}_1, 1) + \sum_{r=2}^{t-1} [f(\mathbf{x}_r, r) - w \| \mathbf{x}_r - \mathbf{x}_{r-1} \| ] + [f(\mathbf{x}, t) - w \| \mathbf{x} - \mathbf{x}_{t-1} \| ] \}
\label{eqn:local_score_function}
\end{equation}

Then $s_{t}(\mathbf{x})$ can be calculated using $s_{t-1}(\cdot)$ as follows:

\begin{equation}
s_{t}(\mathbf{x}) = \max_{\mathbf{y} \in \mathbf{X}} \,\, [s_{t-1}(\mathbf{y}) + f(\mathbf{x}, t) - w \| \mathbf{x} - \mathbf{y} \| ]
\label{eqn:optimal_dependency}
\end{equation}

Thus given $\mathbf{x}$, the calculation of $s_{t}(\mathbf{x})$ will automatically identify $\mathbf{x}_{t-1}$ that would achieve maximum $s_{t}(\mathbf{x})$. In addition, ${\operatorname{arg\,max}}_{\mathbf{x}} \, s_T(\mathbf{x})$ gives the location where the best track terminates at time $T$. This provides foundation of tracking. The procedure of optimal scores and trace-back can be summarized in the following algorithms. Similar approach has been used by Sage et al. ~\cite{sage2005ati} to track a single particle in noisy images.

\begin{algorithm}[th]
\caption{{\tt DPScoring}}
\begin{tabular}{@{\hspace{-0.3ex}}p{0.12\columnwidth}p{0.02\columnwidth}p{0.78\columnwidth}}
\textbf{Input}: & (1) & $\mathbf{X}$, the set of all locations\\
                & (2) & $T$, total number of time points\\
                & (3) & $f(\mathbf{x}, t)$, the intensity function of image sequences of $T$ time points\\
\textbf{Output}:& (1) & $s_t(\mathbf{x})$, the function of maximum score from all tracks that terminates at $\mathbf{x}$ at time $t$\\
				& (2) & $b_t(\mathbf{x})$, the trace back function that indicates the location of track (corresponding to $s_t(\mathbf{x})$) at time $t-1$\\
\end{tabular}

\begin{algorithmic}[1]

\FORALL{$\mathbf{x}$ in $\mathbf{X}$}
\STATE $s_{1}(\mathbf{x}) \leftarrow f(\mathbf{x}, 1)$; \COMMENT{ In discrete space case, $s_{t}(\mathbf{x})$ is an $n$ dimensional array. Its elements are to be determined in this algorithm. Same is for $b_t(\mathbf{x})$.}
\ENDFOR

\FOR{$t=2$ to $T$}
	\FORALL{$\mathbf{x}$ in $\mathbf{X}$}
		\STATE $s_{t}(\mathbf{x}) \leftarrow \max_{\mathbf{y} \in \mathbf{X}} \,\,  [s_{t-1}(\mathbf{y}) + f(\mathbf{x}, t) - w \| \mathbf{x} - \mathbf{y} \| ]$;
		\STATE $b_{t}(\mathbf{x}) \leftarrow {\operatorname{arg\,max}}_{\mathbf{y} \in \mathbf{X}}  \,\, [s_{t-1}(\mathbf{y}) + f(\mathbf{x}, t) - w \| \mathbf{x} - \mathbf{y} \| ]$;
	\ENDFOR
\ENDFOR

\end{algorithmic}
\label{alg:DPScoring}
\end{algorithm}

\begin{algorithm}[th]
\caption{{\tt DPTraceback}}
\begin{tabular}{@{\hspace{-0.3ex}}p{0.12\columnwidth}p{0.02\columnwidth}p{0.78\columnwidth}}
\textbf{Input}: & (1) & $s_t(\mathbf{x})$, the function of maximum score from all tracks that terminates at $\mathbf{x}$ at time $t$\\
				& (2) & $b_t(\mathbf{x})$, the trace back function that indicates the location of track (corresponding to $s_t(\mathbf{x})$) at time $t-1$\\
\textbf{Output}: & \multicolumn{2}{p{0.88\columnwidth}}{$\mathbf{x}_1, \ldots, \mathbf{x}_T$, the locations in the track that achieves maximum score $s_T$}
\end{tabular}

\begin{algorithmic}[1]

\STATE $\mathbf{x}_T \leftarrow {\operatorname{arg\,max}}_{\mathbf{x} \in \mathbf{X}} \,\, s_T(\mathbf{x})$

\FOR{$t=T-1$ to $1$}
\STATE $\mathbf{x}_t \leftarrow b_{t+1}(\mathbf{x}_{t+1})$
\ENDFOR

\end{algorithmic}
\label{alg:DPTraceback}
\end{algorithm}

\section{Experimental results}\label{sec:single_object_result}
\begin{figure} [tph]
\centering
\includegraphics[width=0.32\columnwidth, angle=90] {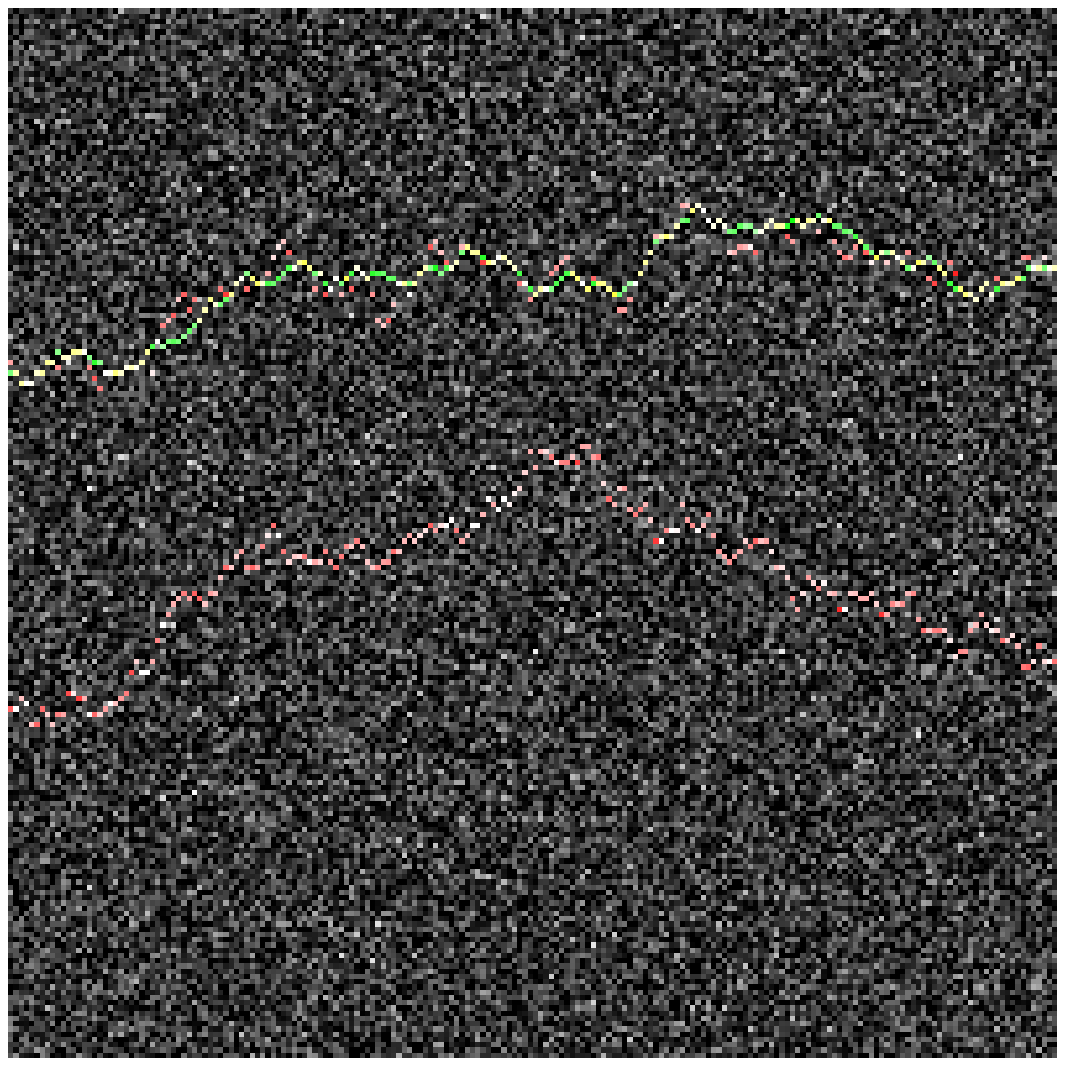}
\includegraphics[width=0.32\columnwidth, angle=90] {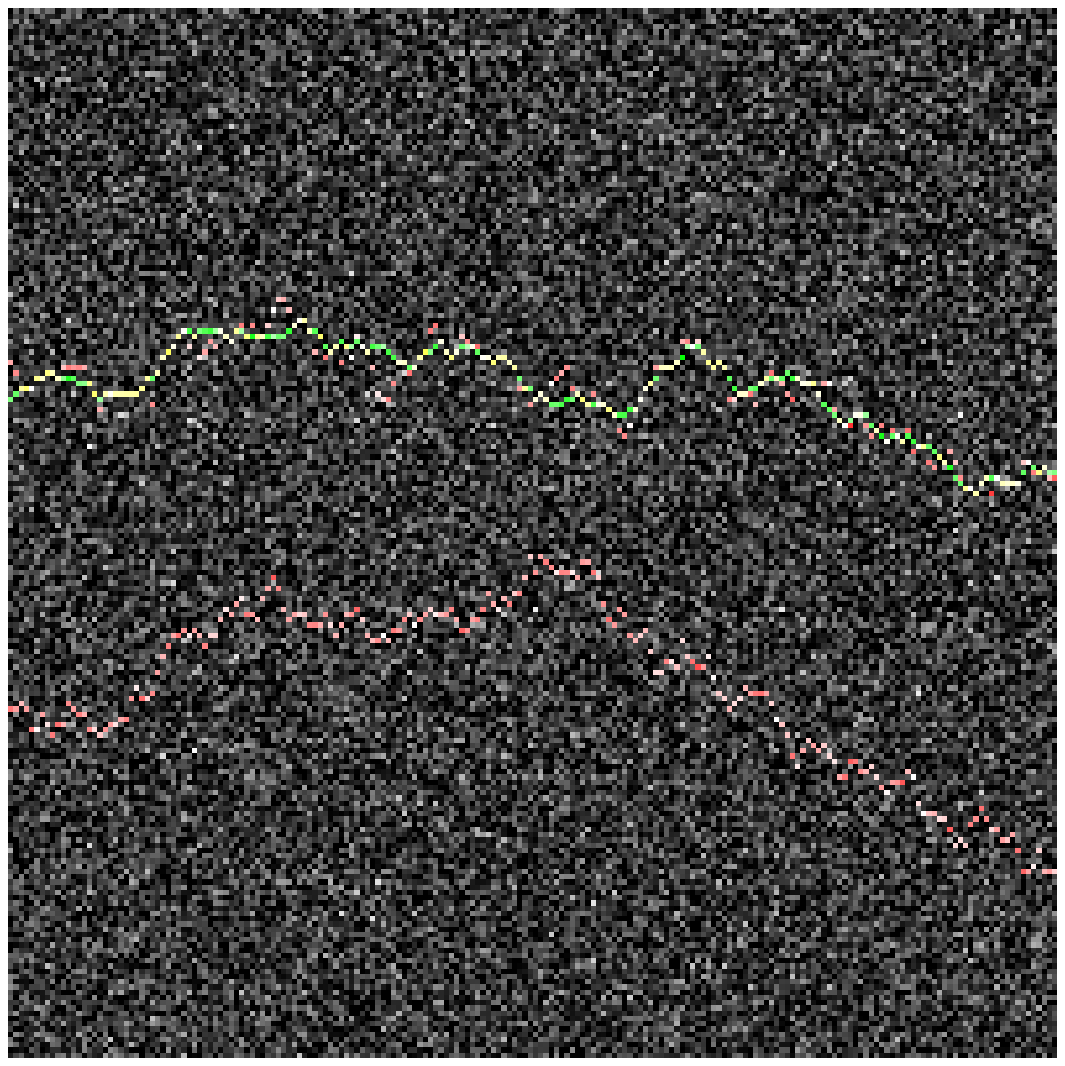}
\includegraphics[width=0.32\columnwidth, angle=90] {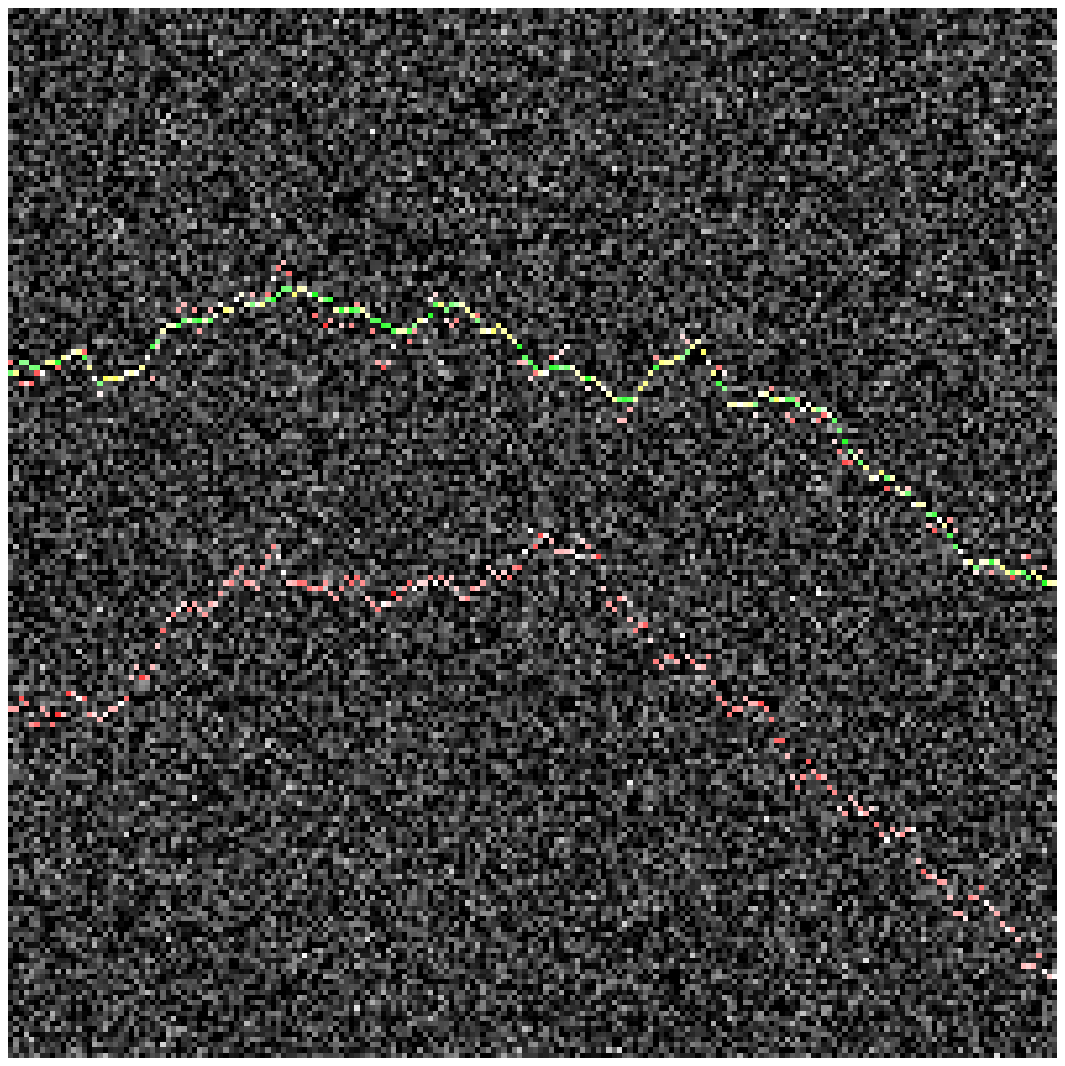}

\caption{Examples of tracking two moving objects in a noisy video.}
\label{fig:single_traceback}
\end{figure}

Figure~\ref{fig:single_traceback} shows the experiments of applying the dynamic programming approach. The left to right sub-figures correspond to experiment 1, 2 and 3 respectively. Two red curves indicate the true positions of the objects. Green curve indicates the inferred positions of a track using the dynamic programming approach. The objects move up and down, and the video proceeds from left to right.

In the first experiment, we assume $n=1$. That is, the object is moving in an one dimensional discrete space. Then each image can be represented as a column vector (of length 200). They are put together from left to right according to order in time. The images contain two objects moving over time of length 200. The objects start from fixed positions, and their movements are modeled as discretized Brownian motion, i.e. a normal distribution $N(0, 2)$. The true intensity of the object is 0.5, and the noise follows a normal distribution of $N(0.2, 0.2)$.

Because the dynamic programming approach does not assume any movement model, it can be used to track more complicated movements. In the second experiment, addition to the first experiment, we further add a constant shift $-0.2$ to the motion, i.e. the motion follows $N(-0.1, 2)$. In the third experiment, we assume there is a constant acceleration $-0.003$, i.e. the motion follows $N(-0.003 t, 2)$. In all the three experiments, the tracking algorithm correctly inferred one of the tracks.

To obtain another track, it is possible to trace back from the position of the second best score. However, in practice, in general we don't know the number of objects. Also, objects may emerge or disappear during image recording.

\subsection{Tracking accuracy}
\textbf{Performance measure:} we use the \emph{Root Mean Squared Error} (RMSE) to measure the tracking accuracy ~\cite{sage2005ati}. This measure is defined as $\operatorname{RMSE}(\hat{\mathbf{x}}) = \sqrt{\sum_t \| \hat{\mathbf{x}}_t - \mathbf{x}_t \| ^2 }$, which is the difference between the estimated track $\hat{\mathbf{x}}$ and true track $\mathbf{x}$. Here the expectation is simply approximated by averaging different realization across time.

\textbf{Average performance on different parameters and configurations: }
Starting from the setting of the experiment 1 in Figure~\ref{fig:single_traceback}, we simulated the data tracking by 1) using different norm (specified by different powers) in the dynamical programming track estimation algorithm, 2) using different weight $w$, 3) adding different amount of noise in the images, 4) different variance in the Brownian motion.

\begin{figure} [tph]
\centering
\includegraphics[width=0.49\columnwidth] {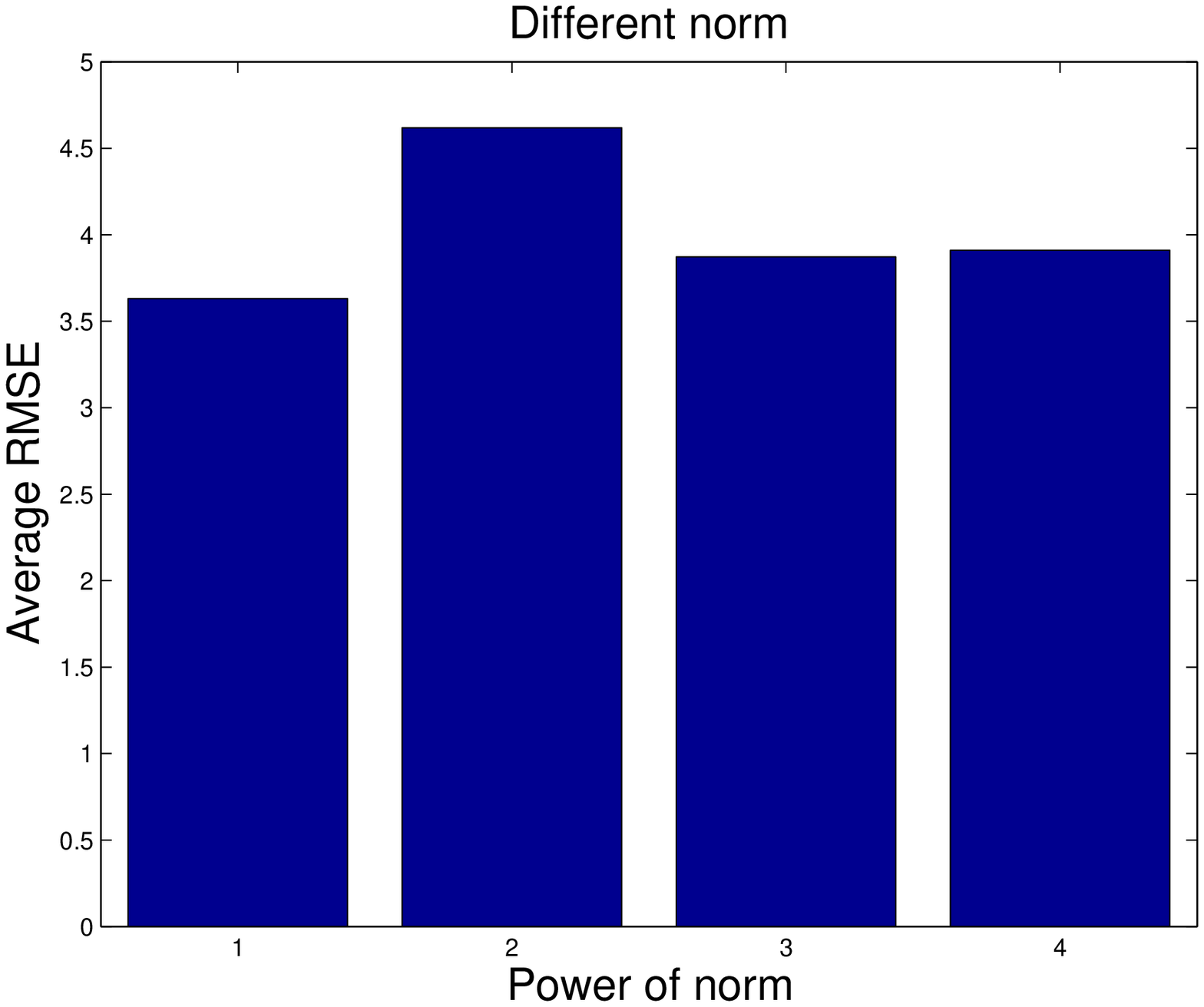}
\includegraphics[width=0.49\columnwidth] {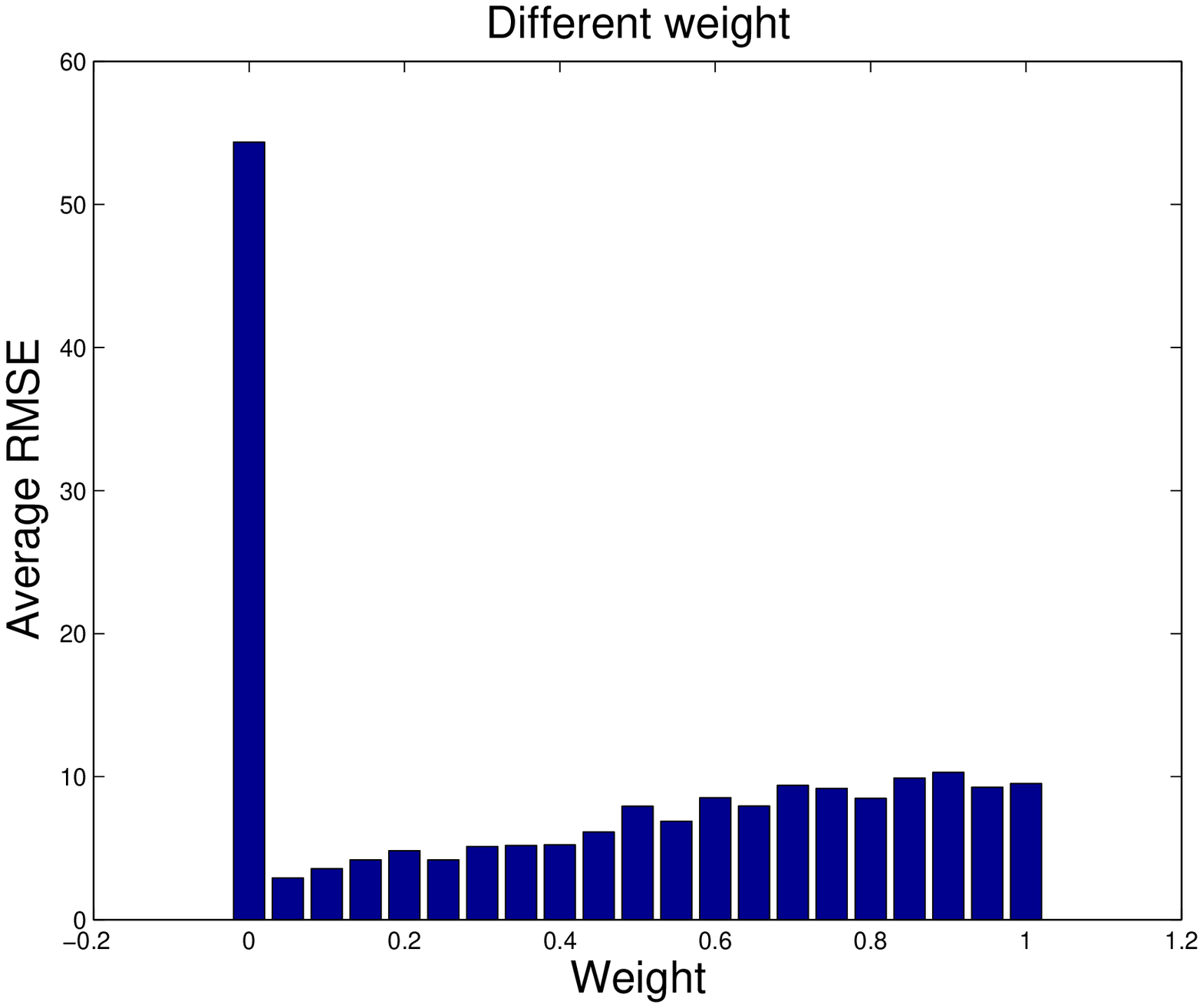}
\includegraphics[width=0.49\columnwidth] {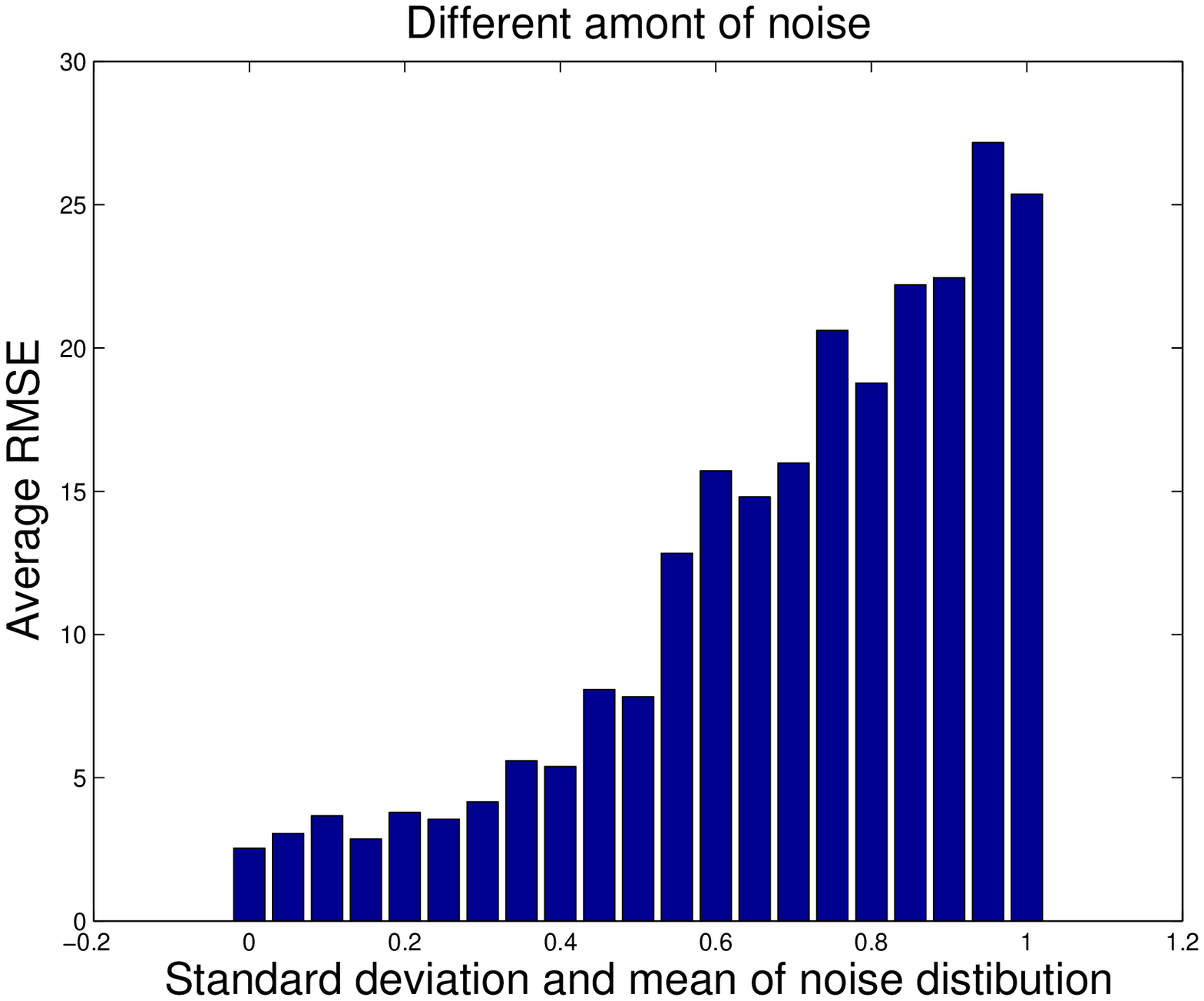}
\includegraphics[width=0.49\columnwidth] {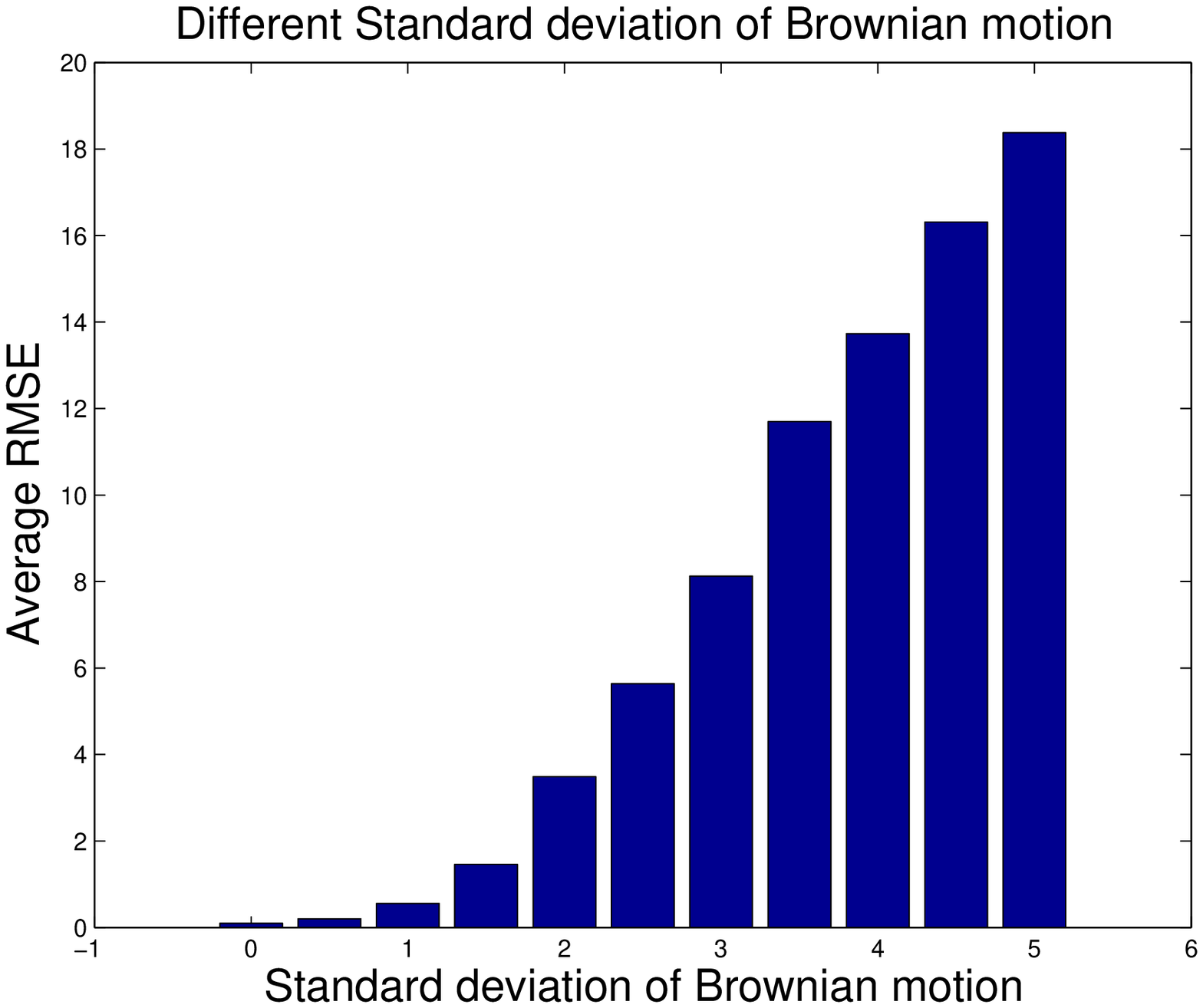}
\caption{Performance analysis of different settings. }
\label{fig:batch_test}
\end{figure}

For each configuration, the simulation and tracking are repeated 100 times and the average RMSE is calculated. Since there are two objects in the model but only one track is estimated, for each simulation, the smaller of the two RMSEs are chosen for averaging. The results are summarized in Figure~\ref{fig:batch_test}. Note that the scales of different plots are different.

It can be seen from the figure, generally, 1) the tracking is insensitive to the choice of norms, 2) the smaller the magnitude of weight, the more accurate the tracking; 3) the tracking performance is similar when the level of noise (indicated by mean and standard deviation of the normal distribution of noise) is less than 0.3, but decreases quickly when noise level increases from 0.3; 4) the faster the movement of objects (indicated by the standard deviation of Brownian motion), the higher the tracking errors.

\textbf{Selection of the optimal track: }
In the experiment 1 of Figure~\ref{fig:single_traceback}, only one of the two tracks is selected. To measure the contribution of noise to the choice of one track against another, we fixed the object tracks as in the experiment, and repeated tracking of the videos with different instances of noise for 1000 times. We found the lower track is selected with a frequency of 39.8\%, the upper one is with frequency 60.2\%.

The reason on why there is a slightly higher chance for the upper track be selected is probably due to the shapes of the two tracks: the upper track has a smaller overall drift than the lower track. In such case, the noise nearby the track, if realized in a high value, can be included into the estimated track to reduce the cost of displacement, therefore generating a higher dynamic programming score.

\begin{figure} [tph]
\centering
\includegraphics[width=0.8\columnwidth, angle=90] {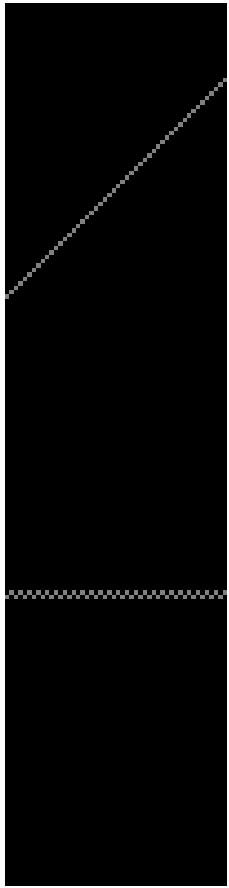}
\caption{Simulated noise-free video that consists of one zigzag track and one linear track.}
\label{fig:zigzag_track}
\end{figure}

To demonstrate this effect, we repeated the above experiment on a new set of videos (of 200 pixels in size and 50 time points) containing two objects. The moving displacements of both objects are 1 between two consecutive frames. But one object keep change moving direction, therefore following a zigzag track. While the other object does not change direction, therefore following a linear track. Figure~\ref{fig:zigzag_track} shows a noise free video that consists these two tracks. Without the presence of noise, the two tracks would result in the same dynamic programming scores. However, the noise provided positive contributions to those tracks with small drift. Our experiment shows the frequency of the two tracks being selected are 68.8\% and 31.2\% respectively.

\chapter{Tracking of multiple vesicles} \label{chp:multi}
\section{Track elimination and enumeration}
The dynamic programming approach can only generate track of one object. To track multiple objects, in practice, one can generate the estimation of multiple tracks by repetitively eliminating the signal along the track inferred by dynamical programming. The following MultiTrack algorithm gives an example of generating multiple tracks.

\begin{algorithm}[th]
\caption{{\tt MultiTrack}}
\begin{tabular}{@{\hspace{-0.3ex}}p{0.12\columnwidth}p{0.02\columnwidth}p{0.78\columnwidth}}
\textbf{Input}: & (1) & $n_{\mathrm{trk}}$, the number of tracks to enumerate\\
				& (2) & $\mathbf{X}$, the set of all locations\\
                & (3) & $T$, total number of time points\\
                & (4) & $f(\mathbf{x}, t)$, the intensity function of image sequences of $T$ time points\\
\textbf{Output}: & \multicolumn{2}{p{0.88\columnwidth}}{$(\mathbf{x}_1^{(1)}, \ldots, \mathbf{x}_T^{(1)}), \ldots, (\mathbf{x}_1^{(n_{\mathrm{trk}})}, \ldots, \mathbf{x}_T^{(n_{\mathrm{trk}})})$, the best tracks}
\end{tabular}

\begin{algorithmic}[1]

\FOR{$i=1$ to $n_{\mathrm{trk}}$}
\STATE $(s, b) \leftarrow \mathrm{DPScoring} (\mathbf{X}, T, f)$
\STATE $(\mathbf{x}_1^{(1)}, \ldots, \mathbf{x}_T^{(1)}) \leftarrow \mathrm{DPTraceback} (s, b)$
\STATE $f \leftarrow \mathrm{TrackElimination} (\mathbf{X}, T, f, (\mathbf{x}_1^{(1)}, \ldots, \mathbf{x}_T^{(1)}))$
\ENDFOR

\end{algorithmic}
\label{alg:MultiTrack}
\end{algorithm}

The MultiTrack calls TrackElimination algorithm to eliminate a track in the video. It is described as follows.

\begin{algorithm}[th]
\caption{{\tt TrackElimination}}
\begin{tabular}{@{\hspace{-0.3ex}}p{0.12\columnwidth}p{0.02\columnwidth}p{0.78\columnwidth}}
\textbf{Input}: & (1) & $\mathbf{X}$, the set of all locations\\
                & (2) & $T$, total number of time points\\
                & (3) & $f(\mathbf{x}, t)$, the intensity function of image sequences of $T$ time points\\
                & (4) & $(\mathbf{x}_1, \ldots, \mathbf{x}_T)$, one track \\
\textbf{Output}: & \multicolumn{2}{p{0.88\columnwidth}}{$f(\mathbf{x}, t)$, the intensity function of image sequences with eliminated track}
\end{tabular}

\begin{algorithmic}[1]

\FOR{$t=1$ to $T$}
\STATE $\mathbf{x}' \leftarrow$ a random location in $\mathbf{X}$
\STATE $f(\mathbf{x_t}, t) \leftarrow f(\mathbf{x}', t)$
\ENDFOR

\end{algorithmic}
\label{alg:TrackElimination}
\end{algorithm}

For simplification, the above algorithm only stops when a fixed number of $n_{\mathrm{trk}}$ is obtained. A more rigorous stopping criterion may be obtained by comparing the scoring function $s_{T}$ obtained from true video against $s_{T}$ from a permutated video through randomly shuffling its pixels.

In addition, the above TrackElimination algorithm assumes the object is only of one pixel. In practice, one object on the track may occupy a small consecutive region instead of just one pixel, we can assume that the track is the trajectory of the center of the object. To remove this object, we can approximate the object using a Gaussian function and subtract the values of this function from the image. To be more specific, denote the gaussian function as $f_t(\mathbf{x}) = a_t \exp(- { (\mathbf{x}-\mathbf{x}_t)^T \mathbf{\Sigma}_t^{-1} (\mathbf{x}-\mathbf{x}_t) })$, where $\mathbf{x}_t$ are obtained from trace back information, but $a_t$ and $\mathbf{\Sigma}_t$ are to be estimated so that $f_t$ would have best least square fit to the image. McKenna et al. used similar idea for the modeling of object intensities for tracking \cite{mckenna1999tco}, where they model the pixel intensities of an object as random variables that follow bivariate Gaussian distribution. They then used expectation maximization to estimate the mean and covariance matrix of the Gaussian distribution, obtaining the best approximation of the object.


\subsection{Experimental results}

\begin{figure} [tph]
\centering
\includegraphics[width=0.45\columnwidth, angle=90] {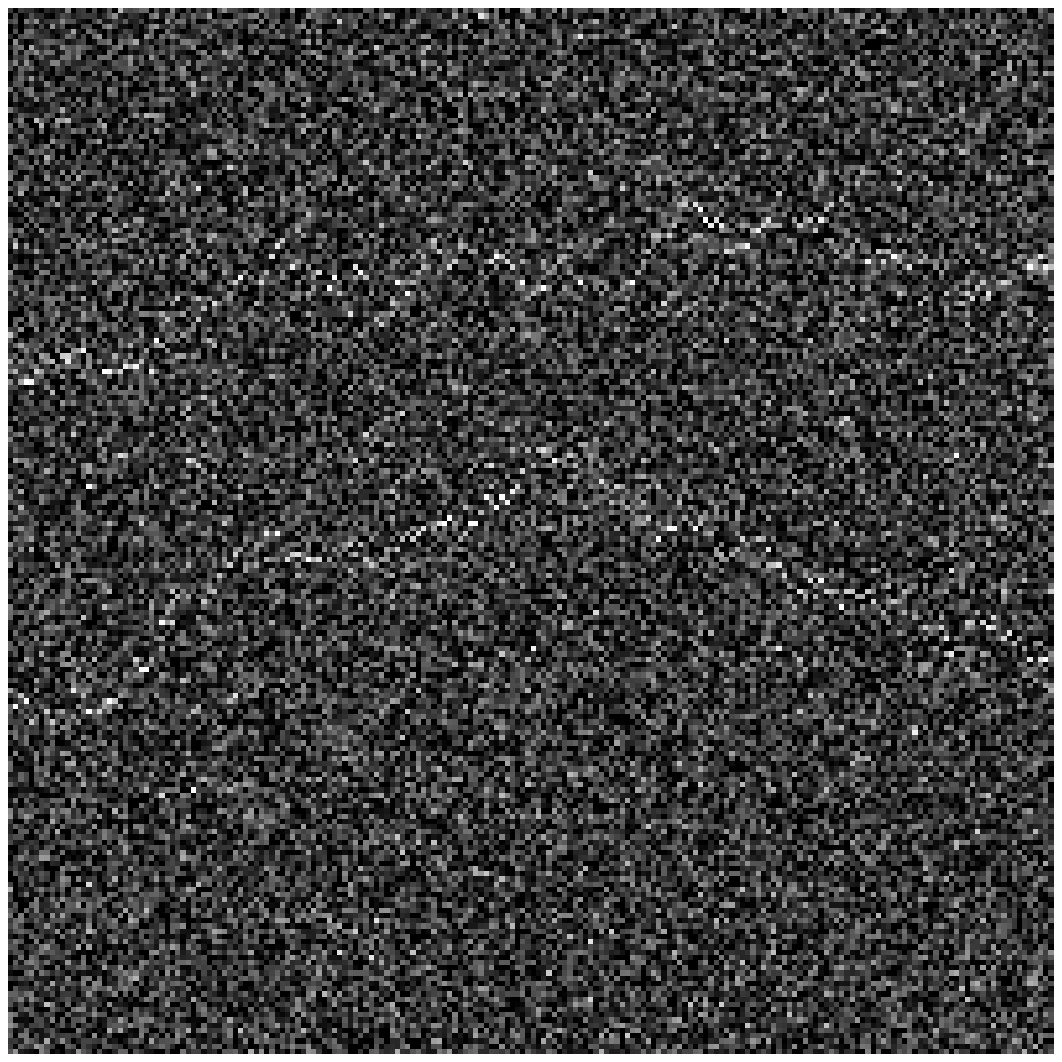}
\includegraphics[width=0.45\columnwidth, angle=90] {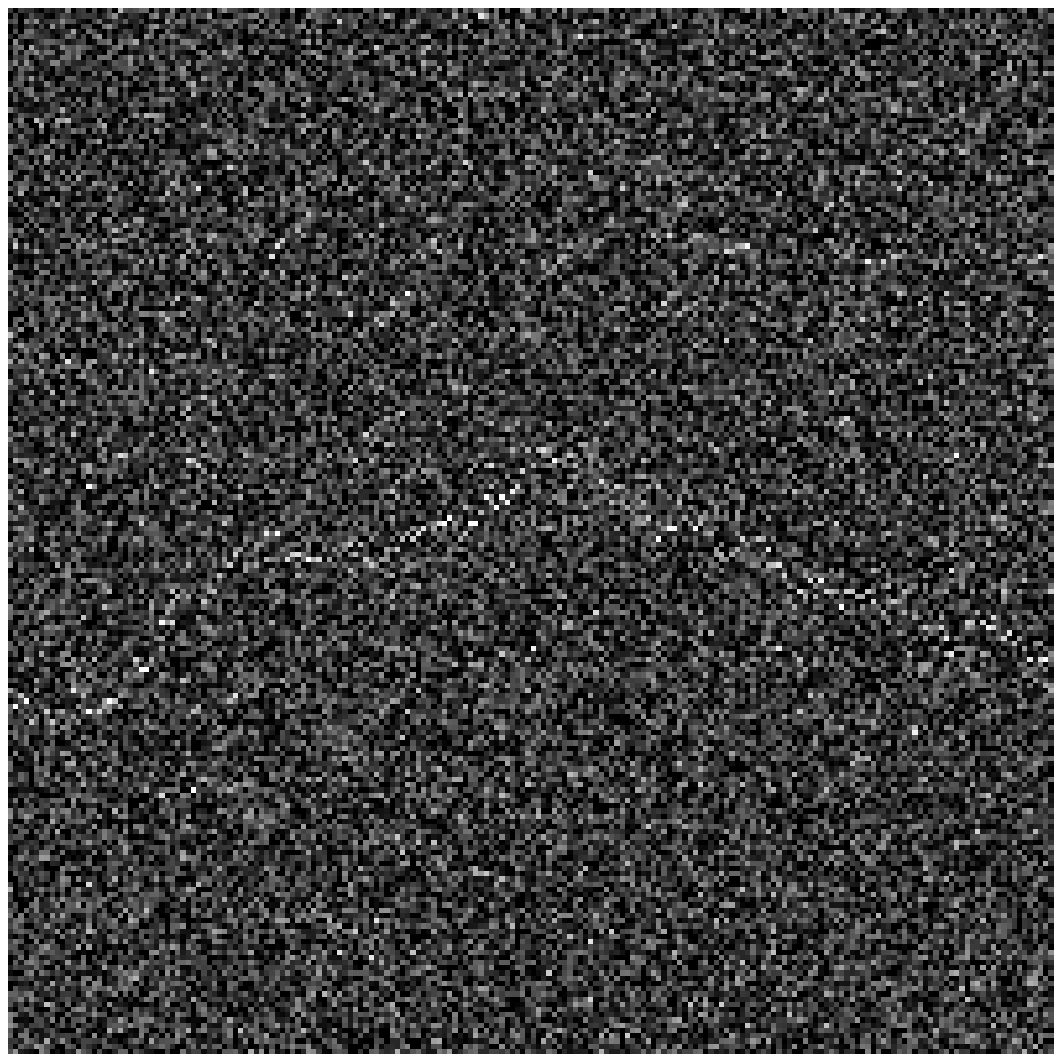}
\caption{Left: Simulated video used in Section ~\ref{sec:single_object_result} (same as Figure ~\ref{fig:single_traceback} without estimated track). Right: the video with the best track eliminated.}
\label{fig:single_traceback_TrackElimination}
\end{figure}

An example of applying TrackElimination to the simulation generated in Section ~\ref{sec:single_object_result} is shown in Figure ~\ref{fig:single_traceback_TrackElimination}.

\textbf{The accuracy of second track after eliminating first track:}
We use the following steps to test the tracking performance of the second object by following steps: 1) simulate videos using the same model as in Section ~\ref{sec:single_object_result}; 2) use algorithms DPScoring and DPTraceback to obtain the track of one object; 3) use TrackElimination to eliminate the track from the video; 4) use DPScoring and DPTraceback to get the second estimated track. We repeated the above steps 1000 times. We find the RMSE of the first tracking from step 3 is $2.29 \pm 0.89$, and the average RMSE of the second tracking from step 4 is $2.30 \pm 0.55$. These two RMSEs are very comparable. We conclude that the track elimination strategy can successfully estimate tracks of multiple objects.

\section{Improving track associations using Kalman filter}
In this section, we propose to improve the tracking accuracy by incorporating Kalman filter into dynamic programming framework.

\subsection{Motivation}
\begin{figure} [tph]
\centering
\includegraphics[width=0.5\columnwidth, angle=90] {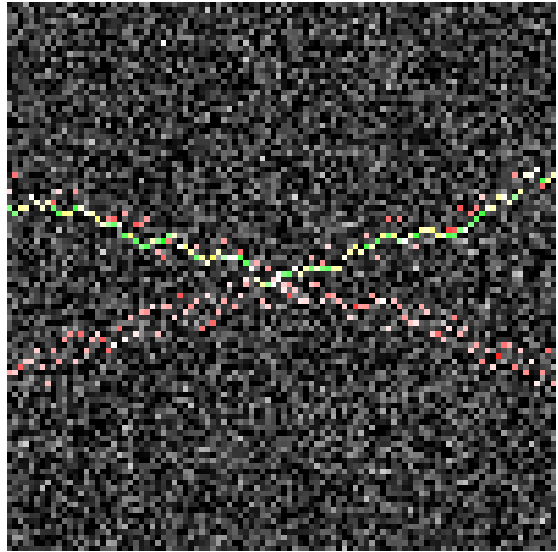}
\caption{Wrong tracking of two objects due to cross of their real tracks.}
\label{fig:single_traceback_cross}
\end{figure}

In practice, when a video contains multiple moving objects, the trajectories of these objects can often get very close or even cross each other. In this case, the previous dynamical programming approach may infer a track that is actually a mixture of multiple real tracks. Figure ~\ref{fig:single_traceback_cross} shows such an example, resulting from a video of 100 pixels and the 100 time points. In this example, the real track (indicated by red lines) of two objects are two cross straight lines plus small random displacements following discritized normal distribution. The green curve corresponds to the estimated track. It can be seen in this figure that the estimated track is actually a combination of first part of one track and the second part of another track. To avoid such problem, we modify the score function in Equation ~\ref{eqn:score_function} to incorporate Kalman filter which provides estimate of current object state.

\subsection{Introduction to Kalman filter}

Kalman filters are based on linear dynamical systems in discrete time domain. Let the state of the system at time $t$ be represented as a real vector $\textbf{s}_{t}$. The Kalman filter model assumes the true state at time $t$ is evolved from the state at $(t - 1)$ according to

\begin{equation}
    \textbf{s}_{t} = \textbf{F}_{t} \textbf{s}_{t-1} + \textbf{B}_{t} \textbf{u}_{t} + \textbf{w}_{t}
\end{equation}

\noindent where
\begin{itemize}
\item $\textbf{F}_{t}$ is the state transition operator applied to the previous state $\textbf{s}_{t-1}$;

\item $\textbf{B}_{t}$ is the control-input operator applied to the external control vector $\textbf{u}_{t}$;

\item $\textbf{w}_{t}$ is the process noise. It is assumed to be sampled from a zero mean multivariate normal distribution with covariance $\textbf{Q}_{t}$.

\end{itemize}

At time $t$ an indirect measurement $\textbf{z}_{t}$ of the true state $\textbf{s}_{t}$ is observed according to
\begin{equation}
    \textbf{z}_{t} = \textbf{H}_{t} \textbf{s}_{t} + \textbf{v}_{t}
\end{equation}

\noindent where $\textbf{H}_{t}$ is the operator that transforms the true state space into the observed space and $\textbf{v}_{t}$ is the observation noise. The noise is assumed to be zero mean Gaussian white noise with covariance $\textbf{R}_{t}$.
\begin{equation}
    \textbf{v}_{t} \sim N(0, \textbf{R}_t)
\end{equation}

\noindent In addition, the initial state, and the noise vectors at each step $\{ \textbf{s}_{0}, \textbf{w}_{1}, ..., \textbf{w}_{t}, \textbf{v}_{1} ... \textbf{v}_{t} \}$ are all assumed to be mutually independent.

The Kalman filter method consists of two phases: predict and update. In the predict phase, the current state estimate is generated using previous state estimates:

\begin{itemize}

\item Predicted state $\hat{\textbf{s}}_{t|t-1} = \textbf{F}_{t}\hat{\textbf{s}}_{t-1|t-1} + \textbf{B}_{t-1} \textbf{u}_{t-1}$

\item Predicted estimate covariance $\textbf{P}_{t|t-1} = \textbf{F}_{t} \textbf{P}_{t-1|t-1} \textbf{F}_{t}^{T} + \textbf{Q}_{t-1}$
\end{itemize}

In the update phase, the currently observed measurement information is used to refine the prediction:

\begin{itemize}

\item Innovation or measurement residual $\tilde{\textbf{y}}_{t} = \textbf{z}_{t} - \textbf{H}_{t}\hat{\textbf{s}}_{t|t-1}$

\item Innovation (or residual) covariance $\textbf{S}_{t} = \textbf{H}_{t}\textbf{P}_{t|t-1} \textbf{H}_{t}^{T}+\textbf{R}_{t}$

\item Optimal Kalman gain $\textbf{K}_{t} = \textbf{P}_{t|t-1}\textbf{H}_{t}^{T}\textbf{S}_{t}^{-1}$

\item Updated state estimate $\hat{\textbf{s}}_{t|t} = \hat{\textbf{s}}_{t|t-1} + \textbf{K}_{t}\tilde{\textbf{y}}_{t}$

\item Updated estimate covariance $\textbf{P}_{t|t} = (I - \textbf{K}_{t} \textbf{H}_{t}) \textbf{P}_{t|t-1}$

\end{itemize}

\subsection{Improving tracking by incorporating Kalman filter }
{\bf General idea: } We incorporate Kalman filter into our dynamic programming framework as follows: Let $\mathbf{z}_t$ be the vector of observed state of an object, which is usually a combination of the object's location, velocity etc. Assume $\mathbf{z}_t$ is observable at all time. Kalman filter can provide an estimation $\hat{\mathbf{z}}_{t|t-1}$ of the object state at time $t$, given observations up to time $t-1$. Suppose $\mathbf{z}_t$ is of length $m$, Equation ~\ref{eqn:score_function_kalman} gives a modified scoring function,

\begin{equation}
s_T = \sum_{t=1}^T f(\mathbf{z}_t, t) - \sum_{i=1}^{m}[ w_i (\sum_{t=2}^T \| z_t^{(i)} - \hat{z}_{t|t-1}^{(i)} \| ) ]
\label{eqn:score_function_kalman}
\end{equation}

\noindent where $z_t^{(i)}$ is the $i$th element of $\mathbf{z}_t$, and $f(\mathbf{z}_t, t)$ only uses the location elements of $z_t$.
\\
\\
{\bf Design of the dynamic model:} We propose a simple design of the dynamics model for the construction of the Kalman filter for video of one dimensional images. In this case, the true location $\mathbf{x}_t$ is a scalar $x_t$. Assume there is no control on the objects, so we have $\textbf{B}_{t} = 0$ and $\textbf{u}_{t} = 0$. Also, assume $\textbf{F}$, $\textbf{H}$, $\textbf{R}$, and $\textbf{Q}$ are time invariant. We define the state vector as location and velocity of a vesicle.

\begin{equation}
\textbf{s}_{t} = \begin{bmatrix} x_t \\ \dot{x}_t \end{bmatrix}
\end{equation}

We assume that between the $t-1$ and $t$ timestep the vesicle undergoes a constant acceleration of $a_t$ that is normally distributed, with mean 0 and standard deviation $\sigma_a$. Assuming object motion follows Newton's laws, we have

\begin{equation}
\textbf{s}_{t} = \textbf{F} \textbf{s}_{t-1} + \textbf{G}a_{t}
\end{equation}

\noindent where

\begin{equation}
\textbf{F} = \begin{bmatrix} 1 & \Delta t \\ 0 & 1 \end{bmatrix}
\end{equation}

\noindent and

\begin{equation}
\textbf{G} = \begin{bmatrix} \frac{\Delta t^{2}}{2} \\ \Delta t \end{bmatrix}
\end{equation}

\noindent with $\Delta t = 1$. We find that

\begin{equation}
 \textbf{Q} = \textrm{cov}(\textbf{G}a) = \textrm{E}[(\textbf{G}a)(\textbf{G}a)^{T}] = \textbf{G} \textrm{E}[a^2] \textbf{G}^{T} = \textbf{G}[\sigma_a^2]\textbf{G}^{T} = \sigma_a^2 \textbf{G}\textbf{G}^{T}
\end{equation}

\noindent At each time step, a noisy measurement of the true position of the vesicle is made. Assume the noise is also normally distributed, with mean 0 and standard deviation $\sigma_z$.

\begin{equation}
    \textbf{z}_{t} = \textbf{H s}_{t} + \textbf{v}_{t}
\end{equation}

\noindent where

\begin{equation}
    \textbf{H} = \begin{bmatrix} 1 & 0 \\ 0 & 0\end{bmatrix}
\end{equation}

\noindent and

\begin{equation}
    \textbf{R} = \textrm{E}[\textbf{v}_t \textbf{v}_t^{T}] = \begin{bmatrix} \sigma_z^2 \end{bmatrix}
\end{equation}

\noindent In dynamic programming, we assume to know the initial starting state of the vesicle with perfect precision, so we initialize

\begin{equation}
    \hat{\textbf{s}}_{0|0} = \begin{bmatrix} x_0 \\ \dot{x}_0 \end{bmatrix}
\end{equation}

\noindent and to tell the filter that we do not know the exact position and speed, we give it a zero covariance matrix:

\begin{equation}
    \textbf{P}_{0|0} = \begin{bmatrix} B & 0 \\ 0 & B \end{bmatrix}
\end{equation}

\noindent with some large number $B$. The filter will then prefer the information from the first measurements over the information already in the model. Given this above dynamic model, we modify the algorithm DPScoring by replacing the displacement with the object position predicted from Kalman filter, as in Equation ~\ref{eqn:score_function_kalman}, where

\begin{equation}
	\hat{\mathbf{z}}_{t|t-1} = \begin{bmatrix} \hat{z}_{t|t-1} \\ 0 \end{bmatrix} = \textbf{H}_{t}\hat{\textbf{s}}_{t|t-1}
\end{equation}

\subsection{Combining dynamic programming and point detector}
We compare the above integrating Kalman filter with dynamic programming approach with the approach that directly uses Kalman filter to associate the object states estimated by object detectors. Since in our case, the protein vesicles are small, we regard them as points. So we use point detector. A simple point detector that is robust to the noise in the images is based on Gaussian filter. Let $G(\mathbf{x}, \Sigma)$ be a Gaussian function with covariance matrix $\Sigma$, the Gaussian filtering is the convolution of an image (at time $t$) and the Gaussian function. $g(\mathbf{x}, t)=\int{f(\mathbf{y}, t)G(\mathbf{x-y}, \Sigma)} d\mathbf{y}$. After applying the Gaussian filter, the noise in the image are reduced. Then the pixels of $g(\mathbf{x}, t)$ with high intensities may correspond to objects.

Given the detected possible locations of a object at time $0$, we can then use greedy approach through Kalman filter to associate the locations to form a track, as shown in the following algorithm.

\begin{algorithm}[th]
\caption{{\tt DectorKalman}}
\begin{tabular}{@{\hspace{-0.3ex}}p{0.12\columnwidth}p{0.02\columnwidth}p{0.78\columnwidth}}
\textbf{Input}: & (1) & $\mathbf{X}$, the set of all locations\\
                & (2) & $T$, total number of time points\\
                & (3) & $g(\mathbf{z}, t)$, the Gaussian filtered intensity function of image sequences of $T$ time points\\
                & (4) & $\textbf{F}, \textbf{G}, \textbf{Q}, \textbf{H}, \textbf{R}$ , the parameters for the dynamic model\\

\textbf{Output}: & \multicolumn{2}{p{0.88\columnwidth}}{$\mathbf{y}_1, \ldots, \mathbf{y}_T$, the locations in the estimated track}

\end{tabular}

\begin{algorithmic}[1]

\STATE $\mathbf{y}_1 \leftarrow {\operatorname{arg\,max}}_{\mathbf{a} \in \mathbf{X}} g(\mathbf{a}, t)$;

\FOR{$t=2$ to $T$}
	\STATE calculate $\hat{\mathbf{z}}_{t|t-1}$ using Kalman filter
	\STATE $\mathbf{y}_t \leftarrow {\operatorname{arg\,max}}_{\mathbf{a} \in \mathbf{X}} \,\,  [g(\mathbf{a}, t) - w \| \hat{\mathbf{z}}_{t|t-1} - \mathbf{a} \| ]$;
\ENDFOR

\end{algorithmic}
\label{alg:DectorKalman}
\end{algorithm}

\subsection{Experimental results}
\begin{figure} [tph]
\centering
\includegraphics[width=0.5\columnwidth, angle=90] {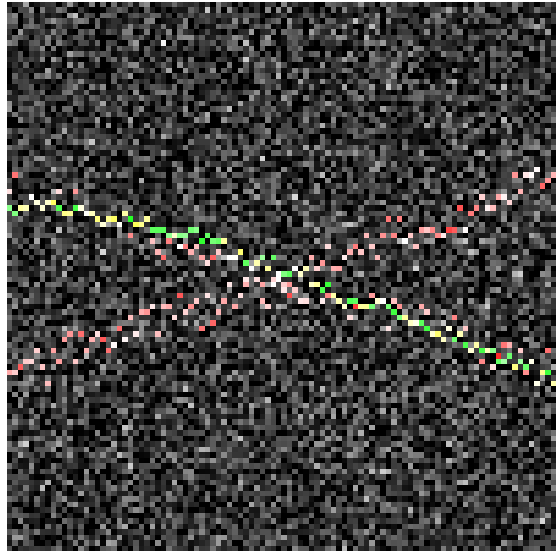}
\caption{Overcoming the wrong association problem by using Kalman filter.}
\label{fig:single_traceback_cross_kalman}
\end{figure}

In our experiment, we assume the protein vesicles are under small amount of acceleration, therefor  we choose $\sigma_a = 0.01$. We assume $\sigma_z = 1$ and $B = 1$. We set $\delta t = 1$. Given these parameters, we tested the above new scoring method on the example data shown in Figure~\ref{fig:single_traceback_cross}. Figure~\ref{fig:single_traceback_cross_kalman} shows that the method correctly finds one of the two tracks.

Given the same true tracks as in the above example, we simulated 1000 noisy videos and compare the performance between tracking using 1) dynamic programming with Kalman filter 2) dynamic programming without using Kalman filter and 3) point detector with Kalman filter (with $\sigma=1$ for Gaussian filtering). The first method gives an RMSE of $2.9 \pm 2.1$, the second method gives RMSE of $8.3 \pm 5.8$ and the third method gives $ 17.6 \pm 14.7 $, suggesting that, compared to the pure dynamic programming approach, the integration of Kalman filter with the dynamic programming greatly reduced the association mistake induced by the cross of two tracks. On the other hand, use Kalman filter alone substantially rely on the accuracy of the point detector. However, in our case, the objects are very small, and the noise is very strong, it is very hard to detect the objects from single images even after filtering. So the third approach resulted in very inaccurate tracking.

\chapter{Conclusion and discussion} \label{chp:discussion}
Automatic tracking of protein vesicle's movements is key to qualitative analysis of the dynamics of these vesicles. The main challenge of such tracking is that the video data is very noisy and the vesicles are very small. In this thesis, after providing an overview of the field of object tracking and their application to the tracking of molecules in cells, we studied the tracking of single and multiple vesicles using dynamic programming and Kalman filter based approaches. Our experiments on simulation data show that dynamic programming approach can achieve high tracking accuracy for single vesicle tracking even there are high levels of noise in the video, and the integration of Kalman filter further significantly increased tracking accuracy by in the case of tracking of multiple vesicles.

Due to the complexity of the vesicle movements, many issues in such tracking remain to be explored. For example, all methods used in this thesis assume the existence of the vesicles in \emph{all} video frames. In real videos, the vesicles could emerge or disappear in some frames. The vesicles may also split or merge. Therefore, more complex association methods like Multiple Hypothesis Testing or Particle Filter may be used to handle this situation.

\begin{singlespace}
    \references[References]{alpha}{main}
\end{singlespace}
\end{document}